\documentclass[12pt, draft, onecolumn, letterpaper]{IEEEtran}

\usepackage{amsmath}
\usepackage{amsfonts, amssymb}
\usepackage{amscd}

\newcommand{\EQ}{\begin{equation}}
\newcommand{\EN}{\end{equation}}

\newtheorem{theo}{Theorem}

\newtheorem{prop}{Proposition}
\newtheorem{lemm}{Lemma}

\newtheorem{ex}{Example}

\newcommand{\pr}{\indent{\em Proof: \ }}
\newcommand{\qed}{\hspace*{5 mm}$\triangle$\bigskip}

\newcommand{\Z}{{\mathbb{Z}}}

\newcommand{\dd}{\displaystyle}

\newcommand{\codi}{{\cal C}}

\newcommand{\Span}{{\cal S}}
\newcommand{\K}{{\cal K}}
\newcommand{\zero}{{\mathbf{0}}}

\newcommand{\add}{\Z_2\Z_4}

\newenvironment{demo}{\noindent {\pr}\ }{\qed}

\newcommand{\cG}{{\cal G}}
\newcommand{\bgamma}{\bar{\gamma}}
\newcommand{\bdelta}{\bar{\delta}}
\newcommand{\bkappa}{\bar{\kappa}}
\newcommand{\RM}{{\mathcal{RM}}}

\title{$\add$-linear codes: rank and kernel
\thanks{This work was supported in part by the Spanish MEC and the
European FEDER under Grants MTM2006-03250 and TSI2006-14005-C02-01. First author wishes to acknowledge the joint sponsorship of the Fulbright Program in Spain and the Ministry of Science and Innovation during a research stay at Auburn University. The material in this paper was presented in part at the XI International Symposium on Problems of Redundancy in Information and Control Systems, Saint Petersburg, Russia, July 2007; and at the 2nd International Castle Meeting on Coding Theory and Applications, Medina del Campo, Spain, September 2008.}}

\author{Cristina Fern\'andez-C\'{o}rdoba, Jaume Pujol and Merc\`{e} Villanueva
\thanks{The authors are members of the Department of Information and Communications Engineering,
                             Universitat Aut\`{o}noma de Barcelona,
                             08193-Bellaterra, Spain.
                             (email:~\{cristina.fernandez, jaume.pujol,
                             merce.villanueva\}@autonoma.edu)}}

\date{\today}

\begin{document}

\maketitle

\begin{abstract}
A code ${\cal C}$ is $\add$-additive if the set of coordinates can
be partitioned into two subsets $X$ and $Y$ such that the punctured code of
${\cal C}$ by deleting the coordinates outside $X$ (respectively, $Y$)
is a binary linear code (respectively, a quaternary linear code).
In this paper, the rank and dimension of the kernel for $\add$-linear codes,
which are the corresponding binary codes of $\add$-additive codes,
are studied. The possible values of
these two parameters for $\add$-linear codes, giving lower and upper
bounds, are established. For each possible rank $r$  between these bounds, the construction of a
$\add$-linear code with rank $r$  is given. Equivalently, for each possible dimension of
the kernel $k$, the construction of a $\add$-linear code with dimension of the kernel
$k$ is given. Finally, the bounds on the rank, once the kernel dimension is fixed, are
established and  the construction of a $\add$-additive code
for each possible pair $(r,k)$ is given.



\end{abstract}

\section{Introduction}

Let $\Z_2$ and $\Z_4$ be the ring of integers modulo 2 and modulo 4,
respectively. Let $\Z_2^n$ be the set of all binary vectors of
length $n$ and let $\Z_4^n$ be the set of all $n$-tuples over the ring $\Z_4$.
In this paper, the elements of $\Z_4^n$ will also be called quaternary vectors of
length $n$.

Any nonempty subset $C$ of $\Z_2^n$ is a binary code
and a subgroup of $\Z_2^n$ is called a {\it binary linear code} or a {\it
$\Z_2$-linear code}. Equivalently, any nonempty subset ${\cal C}$
of $\Z_4^n$ is a quaternary code and a subgroup of $\Z_4^n$ is called
a {\it quaternary linear code}.
Quaternary linear codes can be viewed as binary codes under the usual Gray map defined as
$\phi(0)=(0,0),\ \phi(1)=(0,1), \ \phi(2)=(1,1),\ \phi(3)=(1,0)$ in each coordinate.
If $\codi$ is a quaternary linear code, then the binary code $C=\phi(\codi)$
is called a {\em $\Z_4$-linear} code. The dual of a quaternary linear code $\codi$, denoted by $\codi^{\perp}$,
is called the {\em quaternary dual code} and
is defined in the standard way \cite{MacW} in terms of the usual inner product for quaternary vectors \cite{Sole}.
The binary code $C_{\perp}=\phi(\codi^\perp)$ is called the {\it $\Z_4$-dual code} of $C=\phi(\codi)$.

Since 1994, quaternary linear codes have became significant due to
its relationship to some classical well-known binary codes as the
Nordstrom-Robinson, Kerdock, Preparata, Goethals or Reed-Muller
codes \cite{Sole}. It was proved that the Kerdock code and the
Preparata-like code are $\Z_4$-linear codes and, moreover, the
$\Z_4$-dual code of the Kerdock code is the Preparata-like code.
Lately, more families of quaternary linear codes, called $QRM$,
$ZRM$ and $\RM$, related to the Reed-Muller codes have been studied in
\cite{QRM}, \cite{ZRM} and \cite{PRS08}, respectively.

Additive codes were first defined by Delsarte in 1973 in terms of
association schemes \cite{del}, \cite{lev}. In general, an
additive code, in a translation association scheme, is defined as a
subgroup of the underlying Abelian group.
In the special case of a binary Hamming scheme, that is, when the underlying
Abelian group is of order $2^{n}$, the only structures for the
Abelian group are those of the form $\Z_2^\alpha\times \Z_4^\beta$,
with $\alpha+2\beta=n$. Therefore, the subgroups $\codi$ of
$\Z_2^\alpha\times \Z_4^\beta$ are the only additive codes in a binary Hamming scheme.
In order to distinguish them from additive codes over finite
fields  \cite{Bier}, we will hereafter call them
{\it $\add$-additive codes} \cite{AddDual}, \cite{BR99},  \cite{PheRif}.
The $\add$-additive codes are also included in other families of codes with an algebraic structure,
such as mixed group codes \cite{Lind} and translation invariant propelinear codes  \cite{PropIT}.

Let $\codi$ be a $\add$-additive code, which is a subgroup of
$\Z_2^{\alpha}\times\Z_4^{\beta}$. Let $\Phi: \Z_2^{\alpha}\times\Z_4^{\beta}
\longrightarrow \Z_2^{n}$, where $n=\alpha+2\beta$, be an extension of the
usual Gray map given by
$$\begin{array}{lc}
\Phi(x,y) = (x,\phi(y_1),\ldots,\phi(y_\beta))\\
 \hspace{0,3truecm}\textrm{for any} \,\, x\in\Z_2^\alpha,\; \,\,\, \textrm{and any} \,\, y=(y_1,\ldots,y_\beta)\in \Z_4^\beta.
\end{array}$$
This Gray map is an isometry which transforms Lee
distances defined in a $\add$-additive code $\codi$ over
$\Z_2^{\alpha}\times\Z_4^{\beta}$ to Hamming distances defined in
the corresponding binary code $C=\Phi(\codi)$.  Note that the length of  $C$ is $n=\alpha+2\beta$.

Given a $\add$-additive code $\codi$, the
binary code $C=\Phi(\codi)$ is called a {\em $\add$-linear code}.
Note that $\Z_2\Z_4$-linear codes are a generalization of binary
linear codes and $\Z_4$-linear codes. When $\beta=0$, the binary
code $C=\codi$ corresponds to a binary linear code. On the other
hand, when $\alpha=0$, the $\add$-additive code $\codi$ is a
quaternary linear code and its
corresponding binary code $C=\Phi(\codi)$ is a $\Z_4$-linear code.

Two binary codes $C_1$ and $C_2$ of length $n$ are said to be {\em
isomorphic} if there exists a coordinate  permutation $\pi$ such
that $C_2=\{ \pi(c) \ | \ c\in C_1 \}$. They are said to be {\em
equivalent} if there exists a vector $a\in \Z_2^n$ and a coordinate
permutation $\pi$ such that $C_2=\{ a+\pi(c) \ | \ c\in C_1 \}$. Two
$\add$-additive codes $\codi_1$ and $\codi_2$  are said to be {\it monomially equivalent}, if one can be
obtained from the other by permuting the coordinates and (if
necessary) changing the signs of certain $\Z_4$ coordinates. They are said to be
{\it permutation equivalent} if they differ only by a permutation of
coordinates \cite{Pless}. Note that if two $\add$-additive codes $\codi_1$ and
$\codi_2$ are monomially equivalent, then, after the Gray map,
the corresponding $\add$-linear codes $C_1=\Phi(\codi_1)$ and $C_2=\Phi(\codi_2)$
are isomorphic as binary codes.

Two structural properties of nonlinear binary codes are the rank and
dimension of the kernel. The {\it rank} of a binary code $C$,
$rank(C)$, is simply the dimension of $\langle C \rangle$,
which is the linear span of the codewords
of $C$. The {\em kernel} of a binary code $C$, $K(C)$, is the set of
vectors that leave $C$ invariant under translation, i.e.
$K(C)=\{x\in\Z_2^n \mid C+x=C\}$. If $C$ contains the all-zero
vector, then $K(C)$ is a binary linear subcode of $C$. In general,
$C$ can be written as the union of cosets of $K(C)$, and $K(C)$ is
the largest such linear code for which this is true
\cite{BGH}. We will denote the dimension of the kernel of $C$ by
$ker(C)$.

The rank and dimension of the kernel have been studied for some
families of $\add$-linear codes \cite{QRM}, \cite{PrepKerd}, \cite{ExtPerf},
 \cite{Kro01}, \cite{PePuVi08}, \cite{PhLe95}, \cite{Hadam}. These two
parameters do not always give a full classification of $\add$-linear
codes, since two nonisomorphic $\add$-linear codes could have the
same rank and dimension of the kernel. In spite of that, they can
help in classification, since if two $\add$-linear codes have
different ranks or dimensions of the kernel, they are nonisomorphic.
Moreover, in this case the corresponding $\add$-additive codes are
not monomially equivalent, so these two parameters can also help to distinguish between
$\add$-additive codes that are not monomially equivalent.

Currently, {\sc Magma} supports the basic
facilities for linear codes over integer residue rings and Galois
rings, and for additive codes over a finite field, which are a generalization of the linear codes over
a finite field \cite{M1}. However, it does
not include functions to work with $\add$-additive codes. For this
reason, most of the concepts on $\add$-additive codes
have been implemented recently as a new package in {\sc
Magma}, including the computation of the rank and kernel,
and the construction of some families of $\add$-additive codes \cite{Magma}.

The aim of this paper is the study of the rank and dimension of the kernel of $\add$-linear codes. The paper is organized as follows. In Section \ref{section:preliminaries}, we give some properties related to both $\add$-additive and $\add$-linear codes, including the linearity of $\add$-linear codes. In Section \ref{section:rank}, we determine all possible values of the rank for $\add$-linear codes and we prove the existence of a $\add$-linear code with rank $r$ for all possible values of $r$. Equivalently, in Section \ref{section:kernel}, we establish all possible values of the dimension of the kernel for $\add$-linear codes and we prove the existence of a $\add$-linear code with dimension of the kernel $k$ for all possible values of $k$. In Section \ref{section:pairs}, we determine all possible pairs of values $(r,k)$ for which there exist a $\add$-linear code with rank $r$ and dimension of the kernel $k$ and we construct a $\add$-linear code for any of these possible pairs. Finally, the conclusions are given in Section \ref{section:conclusions}.

\section{Preliminaries}\label{section:preliminaries}

Let $\codi$ be a $\add$-additive code. Since $\codi$ is a subgroup of  $\Z_2^{\alpha}\times \Z_4^{\beta}$,
it is also isomorphic to an Abelian structure
$\Z_2^{\gamma}\times \Z_4^{\delta}$. Therefore, $\codi$ is of type
$2^\gamma4^\delta$ as a group, it has $|\codi|=2^{\gamma+2\delta}$ codewords and
the number of order two codewords in $\codi$ is $2^{\gamma+\delta}$. Let $X$ (respectively $Y$) be the
set of $\Z_2$ (respectively $\Z_4$) coordinate positions, so
$|X|=\alpha$ and $|Y|=\beta$. Unless otherwise stated, the set $X$ corresponds to the first $\alpha$ coordinates and $Y$ corresponds to the last $\beta$ coordinates. Call $\codi_X$ (respectively
$\codi_Y$) the punctured code of $\codi$ by deleting the coordinates
outside $X$ (respectively $Y$). Let $\codi_b$ be the subcode of
$\codi$ which contains all order two codewords and let $\kappa$ be
the dimension of $(\codi_b)_X$, which is a binary linear code. For
the case $\alpha=0$, we will write $\kappa=0$. Considering
all these parameters, we will say that $\codi$ (or equivalently
$C=\Phi(\codi)$) is of type $(\alpha,\beta;\gamma,\delta;\kappa)$.

Although a $\add$-additive code $\codi$ is not a free module,
every codeword is uniquely expressible in the form
$$\dd \sum_{i=1}^{\gamma}\lambda_iu_i+\sum_{j=1}^{\delta}\mu_jv_j,$$ where
$\lambda_i \in \Z_2$ for $1\leq i\leq \gamma$, $\mu_j\in\Z_4$ for
$1\leq j\leq \delta$ and $u_i, v_j$ are vectors in
$\Z_2^\alpha\times \Z_4^\beta$ of order two and four,
respectively. The vectors $u_i,v_j$ give us a generator matrix $\cG$
of size $(\gamma+\delta)\times (\alpha+\beta)$ for the code $\codi$.
Moreover, we can write $\cG$ as \EQ \label{eq:matrixG}
    \cG= \left (\begin{array}{c|c} B_1&2B_3\\ \hline B_2&Q\end{array}\right ),
\EN %
where $B_1,B_2$ are matrices over $\Z_2$ of size $\gamma\times
\alpha$ and $\delta\times \alpha$, respectively;
$B_3$ is a matrix over $\Z_4$ of size $\gamma\times \beta$ with all entries in $\{0,1\}\subset \Z_4$;
and $Q$ is a matrix over $\Z_4$ of size $\delta\times
\beta$ with quaternary row vectors of order four.

Let $I_n$ be the identity matrix of size $n\times n$.
In \cite{Sole}, it was shown that any quaternary linear code of type
$2^\gamma 4^\delta$ is permutation equivalent to a
quaternary linear code with a generator matrix of the form \EQ
\label{eq:QaryStandardForm} \cG_S=\left ( \begin{array}{|ccc} 2T &
2I_{\gamma} & \zero\\ \hline S & R & I_\delta \end{array} \right ),
\EN where $R,T$ are matrices over $\Z_4$ with all entries in $\{0,1\}\subset \Z_4$, and of size
$\delta\times\gamma$ and $\gamma\times(\beta-\gamma-\delta)$,
respectively; and $S$ is a matrix over $\Z_4$ of size
$\delta\times(\beta-\gamma-\delta)$. The following theorem is a
generalization of this result for $\add$-additive codes, so it gives
a canonical generator matrix for these codes.

\begin{theo}\label{prop:StandardForm} \cite{AddDual}
  Let $\codi$ be a $\add$-additive code of type
  $(\alpha,\beta;\gamma,\delta;\kappa)$. Then, $\codi$ is
  permutation equivalent to a $\add$-additive code with canonical generator matrix
  of the form
 \EQ \label{eq:StandardForm}
\cG_S= \left ( \begin{array}{cc|ccc}
I_{\kappa} & T' & 2T_2 & \zero & \zero\\
\zero & \zero & 2T_1 & 2I_{\gamma-\kappa} & \zero\\
\hline \zero & S' & S & R & I_{\delta} \end{array} \right ), \EN
\noindent where $T', S'$ are matrices over $\Z_2$;  $T_1, T_2, R$ are matrices over
$\Z_4$ with all entries in $\{0,1\}\subset \Z_4$;
and $S$ is a matrix over $\Z_4$.
\end{theo}

The concept of duality for $\add$-additive codes was also studied
in \cite{AddDual}, where the appropriate inner product for any two vectors
$u,v\in \Z_2^{\alpha}\times \Z_4^{\beta}$ was defined. Actually, in \cite{AddDual} it was shown that,
given a finite Abelian group, the inner product is uniquely defined
after fixing the generators in each one of the Abelian elementary groups in its decomposition.
In our case, the inner
product in $\Z_2^{\alpha}\times \Z_4^{\beta}$  is
defined over $\Z_4$ as
$$ u \cdot v =2(\sum_{i=1}^{\alpha} u_iv_i)+\sum_{j=\alpha+1}^{\alpha+\beta}
u_jv_j\in \Z_4,$$  where $u,v\in \Z_2^{\alpha}\times
\Z_4^{\beta}$ and the computations are made taking the zeros and ones in the first
$\alpha$ coordinates as quaternary zeros and ones, respectively. If $\alpha=0$, the inner product is the
usual one for quaternary vectors, and if
$\beta=0$, it is twice the usual one for binary vectors.
Then, the {\it additive dual code} of $\codi$, denoted by ${\cal
C}^\perp$, is defined in the standard way $${\cal C}^\perp=\{v\in
\Z_2^\alpha \times \Z_4^\beta \;|\; u \cdot v  =0 \mbox{
for all } u\in {\cal C}\}.$$
The corresponding binary code $\Phi({\cal C}^\perp)$ is denoted by
$C_\perp$ and called the {\it $\add$-dual code} of $C$.
Moreover, in \cite{AddDual} it was proved that the additive dual code $\mathcal{C}^\perp$, which is also a
$\add$-additive code, is of type $(\alpha,\beta;\bgamma,\bdelta;\bkappa)$, where
\EQ \label{dualtype} \begin{array}{l} \bgamma = \alpha + \gamma - 2\kappa,\\
\bdelta =\beta - \gamma - \delta + \kappa,\\
 \bkappa=\alpha-\kappa. \end{array} \EN

The following two lemmas are a generalization of the same results  proved for
quaternary vectors and quaternary linear codes, respectively, in \cite{Sole}. Let $u*v$ denote the component-wise product for any
$u,v \in \Z_2^{\alpha}\times \Z_4^{\beta}$.

\begin{lemm} \label{lem1}
For all $u,v \in \Z_2^{\alpha}\times \Z_4^{\beta}$, we have
$$\Phi(u+v)=\Phi(u)+\Phi(v)+\Phi(2u*v).$$
\end{lemm}

\begin{demo}
Straightforward using the same arguments as for quaternary vectors to prove
that for all $u,v \in \Z_4^\beta$, $\Phi(u+v)=\Phi(u)+\Phi(v)+\Phi(2u*v)$, \cite{Sole}, \cite{Wan}.
\end{demo}

Note that if $u$ or $v$ are vectors in $\Z_2^{\alpha}\times \Z_4^{\beta}$
of order two, then $\Phi(u+v)=\Phi(u)+\Phi(v).$

\begin{lemm} \label{lem2}
Let $\codi$ be a $\add$-additive code.
The $\add$-linear code $C=\Phi(\codi)$ is a binary linear code if and only if
$2u*v\in \codi$ for all $u,v\in \codi$.
\end{lemm}

\begin{demo}
Straightforward by Lemma \ref{lem1} and using the same arguments
as for quaternary linear codes  \cite{Sole}, \cite{Wan}.
\end{demo}

Note that if $\cG$ is a generator matrix of a $\add$-additive code $\codi$ as in (\ref{eq:matrixG}) and  $\{u_i\}_{i=1}^{\gamma}$ and $\{v_j\}_{j=0}^{\delta}$ are the row vectors of order two and
four in $\cG$, respectively, then the $\add$-linear code $C=\Phi(\codi)$ is a binary linear code if and only if
$2v_j*v_k \in \codi$, for all $j, k$ satisfying $1\leq j < k \leq \delta$, since the component-wise product is bilinear.

\section{Rank of $\add$-additive codes}\label{section:rank}

Let $\codi$ be a $\add$-additive code of type $(\alpha,\beta;\gamma,
\delta;\kappa)$  and let $C=\Phi(\codi)$ be the corresponding $\add$-linear code
of binary length $n=\alpha+2\beta$. In this section, we will study the
rank of these $\add$-linear codes $C$. We will
show that there exists a $\add$-linear code $C$ of type
$(\alpha,\beta;\gamma, \delta;\kappa)$ with $r=rank(C)$ for any
possible value of $r$.

\begin{lemm}\label{lemm:RankSet}
 Let $\codi$ be a $\add$-additive code of type $(\alpha,\beta;\gamma,\delta;\kappa)$ and let $C=\Phi(\codi)$ be the corresponding $\add$-linear code. Let $\cG$ be a generator matrix of $\codi$ as in (\ref{eq:matrixG}) and let $\{u_i\}_{i=1}^{\gamma}$ be the rows of order two and $\{v_j\}_{j=0}^{\delta}$ the rows of
order four in $\cG$. Then, $\langle C \rangle$ is generated by
$\{\Phi(u_i)\}_{i=1}^{\gamma}$,
$\{\Phi(v_j),\Phi(2v_j)\}_{j=1}^{\delta}$ and $\{\Phi(2v_j *
v_k)\}_{1\leq j<k\leq \delta}$.
\end{lemm}

\begin{demo}
If $x\in\codi$, then $x$ can be expressed as
$x=v_{j_1}+\dots+v_{j_m}+w$, where
$\{j_1,\dots,j_m\}\subseteq\{1,\dots,\delta\}$ and $w$ is a codeword
of order two. By Lemma \ref{lem1},
$\Phi(x)=\Phi(v_{j_1}+\dots+v_{j_m})+\Phi(w)$, where $\Phi(w)$ is a
linear combination of $\{ \Phi(u_i)\}_{i=1}^{\gamma}$ and
$\{\Phi(2v_j)\}_{j=1}^{\delta}$, and
$\Phi(v_{j_1}+\dots+v_{j_m})=\Phi(v_{j_1})+\dots+\Phi(v_{j_m})+\sum_{1\leq
k<l\leq m}\Phi(2v_{j_k}*v_{j_l})$. Therefore, $\Phi(x)$ is generated by
$\{\Phi(u_i)\}_{i=1}^{\gamma},
\{\Phi(v_j),\Phi(2v_j)\}_{j=1}^{\delta}$ and $\{\Phi(2v_j *
v_k)\}_{1\leq j<k\leq \delta} $.
\end{demo}

\begin{prop}\label{bounds-rank}
Let $C$ be a $\add$-linear code of binary length $n=\alpha+2\beta$
and type $(\alpha,\beta;\gamma, \delta;\kappa)$.  Then,
$$ rank(C)\in \{ \gamma+2\delta,\ldots, \min (\beta+\delta+\kappa, \; \gamma
+2\delta + {\delta \choose 2} ) \}.$$
\end{prop}

\begin{demo}
Let $\cG_S$  be a canonical generator
matrix of $\codi=\Phi^{-1}(C)$ as in (\ref{eq:StandardForm}). In the generator matrix $\cG_S$
there are $\gamma$ rows of order two, $\{u_i \}_{i=1}^\gamma$, and
$\delta$ rows of order four, $\{v_j \}_{j=1}^\delta$. Then, by Lemma \ref{lemm:RankSet},
we can take the matrix $G$ whose row vectors are  $\{\Phi(u_i)\}_{i=1}^{\gamma},
\{\Phi(v_j),\Phi(2v_j)\}_{j=1}^{\delta}$ and $\{\Phi(2v_j *
v_k)\}_{1\leq j<k\leq \delta}$, as a generator matrix of $\langle C \rangle$.

The binary vectors $\{\Phi(u_i)\}_{i=1}^\gamma$ and
$\{\Phi(v_j),\Phi(2v_j)\}_{j=1}^\delta$ are linear independent over
$\Z_2$. Thus, $rank(C)=\gamma+2\delta+\bar{r}$, where $\bar{r}$ is the number of
additional independent vectors taken from $\{\Phi(2v_j*v_k)\}_{1\leq
j<k\leq \delta}$. Note that there are at most ${\delta \choose 2}$ of such vectors. Using row
reduction in $\Phi^{-1}(G)$, the ${\delta \choose 2}$ vectors
$\{ 2v_j*v_k \}_{1\leq j<k\leq \delta} $ can be transformed into vectors with zeroes in the last
$\gamma-\kappa+\delta$ coordinates. Therefore, there are at most
$\min(\beta-(\gamma-\kappa)-\delta, {\delta \choose 2})$ of such
additional independent vectors, so the upper bound of the
rank is $\min(\beta+\delta+\kappa, \; \gamma +2\delta + {\delta
\choose 2})$.

The lower bound follows from the case where the code $C$ is both
binary linear and $\add$-linear.
\end{demo}

Let $\codi$ be a $\add$-additive code of type $(\alpha,\beta;\gamma,\delta;\kappa)$ and let $C=\Phi(\codi)$
with $rank(C)=\gamma+2\delta+\bar{r}$, where $\bar{r}\in\{0,\dots,\min(\beta-(\gamma-\kappa)-\delta,{\delta\choose 2})\}$. Let $\cG$ be a generator matrix of $\codi$ as in (\ref{eq:matrixG}) and let $\{u_i\}_{i=1}^{\gamma}$ be the rows of order two and $\{v_j\}_{j=0}^{\delta}$ the rows of order four in $\cG$. By the proof of Proposition \ref{bounds-rank}, the $\add$-additive code $\Span_{\codi}$ generated by $\{u_i\}_{i=1}^{\gamma}$, $\{v_j \}_{j=1}^{\delta}$ and $\{2v_j *v_k\}_{1\leq j<k\leq \delta}$ is of type $(\alpha,\beta;\gamma+\bar{r},\delta;\kappa)$ and  it is easy to check that $\Phi(\Span_{\codi})=\langle C \rangle$, by Lemma \ref{lemm:RankSet}. Therefore, the code $\langle C \rangle$ is both binary linear and $\add$-linear.

\medskip
For the parameters $\alpha,\beta,\gamma,\delta,\kappa$ given by some
families of $\add$-linear codes such as, for example,
extended 1-perfect $\add$-linear codes (\cite{ExtPerf}, \cite{PheRif} or Example \ref{ex-Z2Z4rank}),
the upper bound above is tight. We
also know $\add$-linear codes such that the rank is in
between these two bounds such as, for example, the Hadamard $\Z_4$-linear codes (\cite{Hadam} or Example \ref{ex-Z2Z4rank}).

\begin{ex} \label{ex-Z4rank} For any integer $t\geq 3$
and each $\delta\in\{1,\ldots,\lfloor (t+1)/2 \rfloor\}$ there exists
a unique (up to isomorphism) extended 1-perfect $\Z_4$-linear code
$C$ of binary length $n=2^t$, such that the $\Z_4$-dual
code of $C$ is of type $(0,\beta;\gamma,\delta)$, where $\beta
=2^{t-1}$ and $\gamma=t+1-2\delta$ \cite{Kro01}. The Hadamard $\Z_4$-linear codes $H$ are the $\Z_4$-dual
of the extended 1-perfect $\Z_4$-linear codes.

The rank of the Hadamard $\Z_4$-linear codes was computed in \cite{Hadam} and the rank
of the extended 1-perfect $\Z_4$-linear codes in \cite{ExtPerf} and \cite{Kro01}.
Specifically,
$$rank(H)=\left\{
 \begin{tabular}{l l l}
   $\gamma+2\delta+{{\delta-1}\choose{2}}$ & if & $\delta \geq 3$ \\
   $\gamma+2\delta$ & if & $\delta=1,2$ \\
\end{tabular}
  \right.$$
and $rank(C)=\bgamma+2\bdelta+\delta=\beta+\bdelta$ (except when $t=4$ and $\delta=1$), where $\bgamma=\gamma$ and $\bdelta=\beta-\gamma-\delta$ by (\ref{dualtype}) taking $\alpha=0=\kappa$. Note that the rank of the extended 1-perfect $\Z_4$-linear codes satisfies the upper bound.
\end{ex}

\begin{ex} \label{ex-Z2Z4rank}
For any integer $t\geq 3$ and each $\delta\in\{0,\ldots,\lfloor
t/2 \rfloor\}$ there exists a
unique (up to isomorphism) extended 1-perfect $\add$-linear code $C$
of binary length $n=2^t$, such that the $\add$-dual code of
$C$ is of type $(\alpha,\beta;\gamma,\delta)$ with $\alpha\not =0$, where $\alpha =
2^{t-\delta}, \beta =2^{t-1}-2^{t-\delta-1}$
and $\gamma=t+1-2\delta$ \cite{BR99}. The Hadamard $\add$-linear codes $H$ are the $\add$-dual
of the extended 1-perfect $\add$-linear codes.

The rank of the Hadamard $\add$-linear codes was computed in \cite{Hadam} and the rank
of the extended 1-perfect $\add$-linear codes in \cite{ExtPerf}. Specifically,
$$rank(H)=\left\{
 \begin{tabular}{l l l}
   $\gamma+2\delta+{{\delta}\choose{2}}$ & if & $\delta \geq 2$ \\
   $\gamma+2\delta$ & if & $\delta=0,1$ \\
\end{tabular}
  \right.$$
and $rank(C)=\bgamma+2\bdelta+\delta=\beta+\bdelta+\bgamma$, where $\bgamma=\alpha-\gamma$ and $\bdelta=\beta-\delta$ by (\ref{dualtype}) taking $\gamma=\kappa$. Note that the rank of these two families of $\add$-linear codes satisfies the upper bound.
\end{ex}

\begin{ex} \label{ex-QRMrank} Let $\overline{QRM}(r,m)$ be the class of $\Z_4$-linear
Reed-Muller codes defined in \cite{QRM}. These are $\add$-linear
codes of type $(0,2^m;0,\delta;0)$, where $\delta=\sum_{i=0}^r
\binom{m}{i}$. An important property is that any $\Z_4$-linear
Kerdock-like code of binary length $4^m$ is in the class
$\overline{QRM}(1,2m-1)$ and any extended $\Z_4$-linear
Preparata-like code of binary length $4^m$ is in the class
$\overline{QRM}(2m-3,2m-1)$.

The rank of any code $C\in \overline{QRM}(r,m)$ is
$$
rank(C)=\sum_{i=0}^r \binom{m}{i} + \sum_{i=0}^t \binom{m}{i},
$$
where $t=\min(2r,m)$, \cite{QRM}. Hence, if $2r\geq m$,
then $rank(C)=\delta + \beta$, i.e. the maximum possible. A
$\Z_4$-linear Kerdock-like code $K$ of binary length $4^m \geq 16$
has $rank(P)=2m^2+m+1$ and an extended $\Z_4$-linear Preparata-like
code $P$ of binary length $4^m \geq 64$ has $rank(P)=2^{2m}-2m$  \cite{PrepKerd}, attaining the upper bound of Proposition
\ref{bounds-rank}.
\end{ex}

The next point to be solved is how to construct $\add$-linear codes with any rank in
the range of possibilities given by Proposition \ref{bounds-rank}.

\begin{lemm}
There exists a $\add$-additive code $\codi$ of type
$(\alpha,\beta;\gamma, \delta;\kappa)$ if and only if
\begin{equation} \label{bounds-code}
 \begin{tabular}{c}
 $\alpha, \beta, \gamma, \delta, \kappa \geq 0$,  $\quad \alpha+\beta >0$, \\
 $0 < \delta+\gamma \leq \beta + \kappa \quad $ \textrm{and} $\quad \kappa
\le \min(\alpha,\gamma)$.
\end{tabular}
\end{equation}
\end{lemm}

\begin{demo}
Straightforward from Theorem \ref{prop:StandardForm}.
\end{demo}

\begin{theo} \label{all-ranks} Let $\alpha,\beta,\gamma,
\delta,\kappa$ be integer numbers satisfying
(\ref{bounds-code}). Then, there exists a $\add$-linear code $C$ of
type $(\alpha,\beta;\gamma, \delta;\kappa)$ with $rank(C)=r$ for any
$$r\in \{ \gamma+2\delta, \ldots, \min (\beta+\delta+\kappa, \; \gamma
+2\delta + {\delta \choose 2} )\}.$$
\end{theo}

\begin{demo} Let $\codi$ be a $\add$-additive code of type
$(\alpha,\beta;\gamma, \delta;\kappa)$ with generator matrix
$$\cG= \left ( \begin{array}{cc|ccc}
I_{\kappa} & T' & \zero & \zero & \zero\\
\zero & \zero & 2T_1 & 2I_{\gamma-\kappa} & \zero\\
\hline \zero & S' & S_r & \zero & I_{\delta} \end{array} \right
),
$$
where $S_r$ is a matrix over $\Z_4$ of size $\delta
\times(\beta-(\gamma-\kappa)-\delta)$, and let $C=\Phi(\codi)$ be
its corresponding $\add$-linear code. Let $\{u_i\}_{i=1}^{\gamma}$ and $\{v_j\}_{j=0}^{\delta}$
be the row vectors of order two and four in $\cG$, respectively.

By Proposition \ref{bounds-rank}, $rank(C)=r=\gamma+2\delta +
\bar{r}$, where $\bar{r} \in \{0, \ldots,
\min(\beta-(\gamma-\kappa)-\delta, {\delta \choose 2}) \}$.
 In the generator matrix $\cG$, the Gray map image of the $\gamma$ row
vectors  $\{u_i\}_{i=1}^\gamma$ and the $2\delta$ row
vectors $\{v_j\}_{j=1}^\delta$,
$\{2v_j\}_{j=1}^\delta$ are independent binary vectors over $\Z_2$.
For each $\bar{r}\in\{0,\ldots,
\min(\beta-(\gamma-\kappa)-\delta, {\delta \choose 2}) \}$,
we will define $S_r$ in an appropriate way such that
$rank(C)=r=\gamma+2\delta+\bar{r}$.

Let $e_k$, $1\leq k \leq \delta$, denote the column vector of length
$\delta$, with a one in the $k$th coordinate and zeroes elsewhere. For
each $\bar{r} \in \{ 0, \ldots ,
\min(\beta-(\gamma-\kappa)-\delta, {\delta \choose 2}) \}$, we can
construct $S_r$ as a quaternary matrix where in $\bar{r}$
columns there are $\bar{r}$ different column vectors $e_k+e_l$ of
length $\delta$, $1\leq k < l\leq \delta$, and in the remaining
columns there is the all-zero column vector. For each one of the
$\bar{r}$ column vectors the rank increases by 1. In fact,
if the column vector $e_k+e_l$ is included in $S_r$, then the quaternary
vector $2v_k*v_l$ has only a two in the same coordinate where the column
vector $e_k+e_l$ is and $\Phi(2v_k*v_l)$ is independent to the vectors
$\{\Phi(u_i)\}_{i=1}^\gamma$,$\{\Phi(v_j)\}_{j=1}^\delta$,
$\{\Phi(2v_j)\}_{j=1}^\delta$ and $\{\Phi(2v_s*v_t)\}$, $\{s,t\}\neq
\{k,l\}$.
Since the maximum number of columns of $S_r$ is
$\beta-(\gamma-\kappa)-\delta$ and the maximum number of different
such columns is $\delta \choose 2$, the result follows.
\end{demo}

Let $S_r$ be a matrix over $\Z_4$ of size $\delta
\times(\beta-(\gamma-\kappa)-\delta)$ where in $\bar{r}=r-(\gamma+2\delta)$ columns there are $\bar r$ different column vectors $e_k+e_l$ of length $\delta$, $1\leq k < l\leq \delta$, and in the remaining columns there are the all-zero column vector. Note that by the proof of Theorem \ref{all-ranks}, any
$\add$-additive code $\codi$ of type $(\alpha,\beta;\gamma,
\delta;\kappa)$ with generator matrix $$\cG= \left (
\begin{array}{cc|ccc}
I_{\kappa} & T' & \zero & \zero & \zero\\
\zero & \zero & 2T_1 & 2I_{\gamma-\kappa} & \zero\\
\hline \zero & S' & S_r & \zero & I_{\delta} \end{array} \right ),
$$
where $T'$, $T_1$ and $S'$ are any matrices over $\Z_2$,
has $rank(\Phi(\codi))=r=\gamma+2\delta +\bar{r}$.

\begin{ex}
By Proposition \ref{bounds-rank}, we know that the possible ranks for
$\add$-linear codes, $C$, of type $(\alpha,9;2,5;1)$ are $rank(C)=r \in \{ 12, 13,14,15\}$.
For each possible $r$, we can construct a $\add$-linear code $C$
with $rank(C)=r$, taking the following generator matrix of $\codi=\Phi^{-1}(C)$:
$$\cG_S= \left ( \begin{array}{cc|ccc}
1 & T'    & \zero & 0 & \zero\\
0 & \zero  & 2T_1 & 2 & \zero\\\hline
\zero& S' &  S_r & \zero & I_5 \end{array} \right ),$$
where
$S_{12}=( \zero )$ and $S_{13}$, $S_{14}$, and $S_{15}$ are constructed as follows:
$$
S_{13}=\left(
    \begin{array}{ccc}
      1 & 0 & 0 \\
      1 & 0 & 0 \\
      0 & 0 & 0 \\
      0 & 0 & 0 \\
      0 & 0 & 0 \\
    \end{array}
  \right),
 \quad
S_{14}=\left(
    \begin{array}{ccc}
      1 & 0 & 0 \\
      1 & 1 & 0 \\
      0 & 1 & 0 \\
      0 & 0 & 0 \\
      0 & 0 & 0 \\
    \end{array}
  \right), \quad
S_{15} =\left(
    \begin{array}{ccc}
      1 & 0 & 1 \\
      1 & 1 & 0 \\
      0 & 1 & 1 \\
      0 & 0 & 0 \\
      0 & 0 & 0 \\
    \end{array}
  \right). $$
\end{ex}

\section{Kernel dimension of $\add$-additive codes}\label{section:kernel}

In this section, we will study the dimension of the kernel of
$\add$-linear codes $C=\Phi(\codi)$. We will also show that there
exists a $\add$-linear code $C$ of type $(\alpha,\beta;\gamma,
\delta;\kappa)$ with $k=ker(C)$ for any possible value of $k$.

\begin{lemm} \label{lem3}
Let $\codi$ be a $\add$-additive code and let $C=\Phi(\codi)$ be the corresponding $\add$-linear code.
Then, $$K(C)=\{\Phi(u) \mid u\in\codi \textnormal{ and } 2u*v\in \codi, \forall v \in\codi\}.$$
\end{lemm}

\begin{demo}
By Lemma \ref{lem2}, $\Phi(u)+\Phi(v)\in C$ if and only if $2u*v \in \codi$ for all $u,v\in\codi$.
Thus, the result follows.
\end{demo}

Note that if $\cG$ is a generator matrix of a $\add$-additive code $\codi$ and $C=\Phi(\codi)$,
$\Phi(u) \in K(C)$ if and only if $u\in \codi$ and $2u*v\in \codi$ for all $v\in \cG$. Moreover, all
codewords of order two in $\codi$ belong to $K(C)$.

\begin{lemm}\label{lemma:Ksubcode}
 Let $\codi$ be a $\add$-additive code and let $C=\Phi(\codi)$ be the corresponding
 $\add$-linear code. Given $x,y\in \codi$, $\Phi(x)+\Phi(y)\in K(C)$ if and only if
 $\Phi(x+y)\in K(C)$.
\end{lemm}

\begin{demo}
By Lemma \ref{lem1}, $\Phi(x+y+2x*y)=\Phi(x)+ \Phi(y)$. Now, by
Lemma \ref{lem3}, $\Phi(x+y+2x*y)\in K(C)$ if and only if for all $v
\in\codi$, $2(x+y+2x* y)*v=2(x+y)* v \in\codi$; that is, if and only
if $\Phi(x+y)\in K(C)$.
\end{demo}

\begin{lemm}\label{prop:all-k}
Let $C$ be a $\add$-linear code of binary length $n=\alpha+2\beta$
and type $(\alpha,\beta;\gamma, \delta; \kappa)$. Then,
$ker(C) \in \{ \gamma +\delta, \gamma+\delta+1, \ldots, \gamma+2\delta-2,\gamma+2\delta \}.$
\end{lemm}

\begin{demo}
The upper bound $\gamma+2\delta$ comes from the linear case. The
lower bound $\gamma+\delta$ is straightforward, since there are
$2^{\gamma+\delta}$ codewords of order two in $\codi=\Phi^{-1}(C)$
and, by Lemma \ref{lem3}, the binary images by $\Phi$ of all these
codewords are in $K(C)$. Also note that if the $\add$-linear code
$C$ is not linear, then the dimension of the kernel is equal to or
less than $\gamma+2\delta-2$ \cite{PhLe95}.
Therefore, $ker(C) \in \{ \gamma +\delta, \ldots,
\gamma+2\delta-2,\gamma+2\delta \}$.
\end{demo}

Given an integer $m>0$, a set of vectors
$\{v_1,v_2,\ldots,v_m\}$ in $\Z_2^{\alpha}\times \Z_4^{\beta}$ and a
subset $I=\{i_1,\ldots,i_l\} \subseteq \{1,\ldots,m\}$, we denote by $v_I$ the vector
$v_{i_1}+\cdots+v_{i_l}$. If
$I=\emptyset$, then $v_I=\zero$. Note that given $I,J\subseteq
\{1,\ldots,m\}$,  $v_I+v_J=v_{(I\cup J)-(I\cap J)} + 2v_{I\cap J}$.

\begin{prop} \label{lemm:cosetsKernel}
Let $\codi$ be a $\add$-additive code of type $(\alpha,\beta;\gamma, \delta; \kappa)$, with generator matrix  $\cG$,
and let $C=\Phi(\codi)$ be the corresponding $\add$-linear code with
$ker(C)=\gamma+2\delta-{\bar k}$, where $\bar k \in \{2,\ldots,\delta\}$.
Then, there exist
a set $\{v_1, v_2, \ldots, v_{\bar k}\}$ of row vectors  of
order four in $\cG$, such that
$$
C= \bigcup_{I \subseteq \{1,\ldots,{\bar k} \}} (K(C)+ \Phi(v_{I}))
$$
\end{prop}

\begin{demo}
We know that $C$ can be written as the union of cosets of $K(C)$
\cite{BGH}.  Since $|K(C)|=2^{\gamma+2\delta-\bar k}$ and
$|C|=2^{\gamma+2\delta}$, there are exactly $2^{\bar k}$ cosets.

Let $u_1,\ldots, u_\gamma, v_1,\ldots,v_\delta $ be the $\gamma$ and
$\delta$ row vectors in $\cG$ of order two and four, respectively.
By Lemma \ref{lem3}, the binary images by $\Phi$ of all codewords of
order two are in $K(C)$. There are $2^{\gamma+\delta}$ codewords of
order two generated by $\gamma+\delta$ codewords. Moreover, there
are $\delta -\bar k$ codewords $w_i$ of order four such that
$\Phi(w_i)\in K(C)$ for all $i\in \{1,\ldots,\delta -\bar k \}$,
and $\Phi(u_1), \ldots, \Phi(u_\gamma),$
$\Phi(2v_1),\ldots,\Phi(2v_\delta),$ $\Phi(w_1),
\ldots,\Phi(w_{\delta-\bar k})$ are linear independent vectors over
$\Z_2$. The code $\codi$ can also be generated by $u_1,\ldots, u_\gamma,
w_1,\ldots, w_{\delta-\bar k}, v_{i_1},\ldots,v_{i_{\bar k}}$, where
$\{ i_1, i_2,\ldots, i_{\bar k} \} \subseteq \{1,\ldots,\delta\}$.
We can assume that  $v_{i_1},\ldots,v_{i_{\bar k}}$ are the $\bar k$
row vectors $v_{1},\ldots,v_{\bar k}$ in $\cG$. Note that
$\Phi(v_I)\not \in K(C)$, for any $I\subseteq \{1,\ldots,{\bar k}\}$
such that $I\neq\emptyset$. In
fact, if $\Phi(v_I) \in K(C)$, then the set of vectors $\Phi(u_1),
\ldots, \Phi(u_\gamma),$ $\Phi(2v_1),\ldots,\Phi(2v_\delta),$
$\Phi(w_1), \ldots,\Phi(w_{\delta-\bar k})$, $\Phi(v_I)$ would be
linear independent.

Finally, we show that the  $2^{\bar k}-1$ binary vectors
$\Phi(v_I)$, $I\subseteq \{1,\ldots,{\bar k}\}$ and $I\neq\emptyset$, are in different cosets. Let
$\Phi(v_I)$ and $\Phi(v_J)$ be any two of these binary vectors such that $I\neq J$. If
$\Phi(v_I) \in K(C)+\Phi(v_J)$, then $\Phi(v_I)+\Phi(v_J) \in K(C)$
and, by Lemma \ref{lemma:Ksubcode}, $\Phi(v_I+v_J) \in K(C)$. We
also have that $v_I+v_J=v_{(I\cup J)-(I\cap J)} + 2v_{I\cap J}$.
Hence, $\Phi(v_{(I\cup J)-(I\cap J)} + 2v_{I\cap J})=\Phi(v_{(I\cup
J)-(I\cap J)})+\Phi(2v_{I\cap J})\in K(C)$ and $\Phi(v_{(I\cup
J)-(I\cap J)})\in K(C)$, which is a contradiction, since $(I\cup
J)-(I\cap J) \subseteq \{1,\ldots,{\bar k}\}$ and $(I\cup
J)-(I\cap J) \neq \emptyset$.
\end{demo}

It is important to note that if $C$ is a $\add$-linear code, then
$K(C)$ is a $\add$-linear subcode of $C$, by Lemma
\ref{lemma:Ksubcode}. The {\it kernel} of a $\add$-additive code
$\codi$ of type $(\alpha,\beta;\gamma, \delta; \kappa)$, denoted by
$\K(\codi)$, can be defined as $\K(\codi)=\Phi^{-1}(K(C))$, where
$C=\Phi(\codi)$ is the corresponding $\add$-linear code. By Lemma
\ref{lem3}, $\K(\codi)=\{ u\in\codi \mid  2u*v\in \codi, \forall v
\in\codi\}$ and it is easy to see that $\K(\codi)$ is a
$\add$-additive subcode of $\codi$ of type $(\alpha,\beta;\gamma
+\bar k, \delta-\bar k; \kappa)$.

Note that replacing ones with twos in the first $\alpha$ coordinates, we can see
$\Z_2\Z_4$-additive codes as quaternary linear codes. Let $\chi$ be
the map from $\Z_2$ to $\Z_4$, which is the usual inclusion from the
additive structure in $\Z_2$ to $\Z_4$: $\chi(0) = 0$, $\chi(1) =
2$. This map can be extended to the map $(\chi, Id) : \Z_2^\alpha
\times \Z_4^{\beta} \rightarrow \Z_4^{\alpha+\beta}$, which will
also be denoted by $\chi$. If $\codi$ is a $\Z_2\Z_4$-additive code of
type $(\alpha,\beta;\gamma,\delta;\kappa)$
with generator matrix $\mathcal{G}$, then $\chi(\codi)$ is a quaternary
linear code of length $\alpha+\beta$ and type $2^\gamma 4^\delta$ with generator matrix
$\mathcal{G}_{\chi(\codi)}=\chi(\mathcal{G})$. Note that
$\K(\codi)=\chi^{-1}\K(\chi(\codi))$ and $\K(\chi(\codi))^\perp$ is
the quaternary linear code generated by the matrix
$$
 \left ( \begin{array}{c}
        \mathcal{H}_{\chi(\codi)} \\
        2\mathcal{G}_{\chi(\codi)} * \mathcal{H}_{\chi(\codi)}\\
 \end{array} \right ),
$$
where $\mathcal{H}_{\chi(\codi)}$ is the generator matrix of the
quaternary dual code of $\chi(\codi)$ and
$2\mathcal{G}_{\chi(\codi)}* \mathcal{H}_{\chi(\codi)}$ is the matrix obtained
computing the component-wise product $2u*v$ for all $u\in
\mathcal{G}_{\chi(\codi)}$, $v\in \mathcal{H}_{\chi(\codi)}$.

Moreover, by Proposition \ref{lemm:cosetsKernel}, given a
$\add$-additive code $\codi$ with generator matrix $\cG$, there
exist a set $\{v_1, v_2, \ldots, v_{\bar k}\}$ of row vectors of
order four in $\cG$, such that
$$
\codi= \bigcup_{I \subseteq \{1,\ldots,{\bar k} \}} (\K(\codi)+ v_{I}).
$$

\begin{lemm}\label{lemm:symmetric-matrix}
Let $A$ be a symmetric matrix over $\Z_2$ of odd order and with
zeroes in the main diagonal. Then, $\det(A)=0$.
\end{lemm}

\begin{demo}
Let $n$ be the order of the matrix $A$. The map $f:\Z_2^n\times\Z_2^n \rightarrow \Z_2^n$ defined by
$f(u,v)=uAv^t$ is an alternating bilinear form and $A$ is a
symplectic matrix \cite[pp. 435]{MacW}. It is known that the
rank $r$ of a symplectic matrix is always even \cite[pp. 436]{MacW}.
Therefore, since the order $n$ of $A$ is an odd number, $r<n$ and $\det(A)=0$.
\end{demo}

\begin{prop}\label{bounds-kernel}
Let $C$ be a $\add$-linear code of binary length $n=\alpha+2\beta$
and type $(\alpha,\beta;\gamma, \delta; \kappa)$ and
$s=\beta-(\gamma-\kappa)-\delta$. Then,
$$\left\{
\begin{tabular}{l}
   if $s=0$, $\quad ker(C)=\gamma +2\delta$, \\
   if $s=1$, $\quad ker(C) \in \{ \gamma +2(\delta-\lceil \frac{\delta-1}{2} \rceil), \ldots, \gamma+2(\delta-1),
  \gamma+2\delta \}$,\\
   if $s\geq 2$, $\quad ker(C) \in \{ \gamma +\delta, \gamma+\delta+1, \ldots, \gamma+2\delta-2,\gamma+2\delta \}.$
\end{tabular} \right.$$
\end{prop}

\begin{demo}
For $s=0$, by Proposition \ref{bounds-rank} we have that $rank(C)=\gamma+2\delta$, so $C$ is a binary linear code and $ker(C)=\gamma+2\delta$. For $s\geq 2$, by Lemma \ref{prop:all-k} we have that $ker(C) \in \{ \gamma +\delta, \ldots, \gamma+2\delta-2,\gamma+2\delta \}$.

Now, we will prove the result for $s=1$.  By Theorem
\ref{prop:StandardForm}, $\codi$ is permutation equivalent to a
$\add$-additive code generated by
$$
\cG_S= \left (
\begin{array}{cc|ccc}
I_{\kappa} & T' & 2T_2 & \zero & \zero\\
\zero & \zero & 2T_1 & 2I_{\gamma-\kappa} & \zero\\
\hline
\zero & S' & S  & R & I_{\delta} \end{array} \right ),
$$
 where $S$ is a
matrix over $\Z_4$ of size $\delta \times 1$.  Let
$\{u_i\}_{i=1}^\gamma$  and  $\{v_j\}_{j=1}^\delta$ be the row
vectors in $\cG_S$ of order two and four, respectively.

If $\delta<3$, then it is easy to see that $ker(C)=\gamma+2\delta-2$
or $ker(C)=\gamma+2\delta$, by Lemma~\ref{prop:all-k}.
If $\delta\geq 3$ we will show that, given four vectors
$v_{j_1},v_{j_2},v_{j_3},v_{j_4}$ such that $2v_{j_1}*v_{j_2}\not\in
\codi$ and $2v_{j_3}*v_{j_4}\not\in\codi$, then $2v_{j_1}*v_{j_2} +
2v_{j_3}*v_{j_4}\in\codi$. Let $e_k$, $1\leq k\leq \alpha+\beta$,
denote the row vector of length $\alpha+\beta$, with a one in the
$k$th coordinate and zeroes elsewhere. Then, we can write
$2v_{j_1}*v_{j_2}=(\zero,\zero,2c,2e_I,\zero)$, where $c\in\{0,1\}$
and $I\subseteq \{\alpha+2,\ldots,\alpha+\gamma-\kappa+1\}$, and
$2v_{j_3}*v_{j_4}=(\zero,\zero,2c',2e_J,\zero)$, where
$c'\in\{0,1\}$ and $J\subseteq
\{\alpha+2,\ldots,\alpha+\gamma-\kappa+1\}$. We denote by $u_I$
(resp. $u_J$) the row vector obtained by adding the row vectors of
order two in $\cG_S$ with $2$ in the coordinate positions given by $I$ (resp.
$J$). Then, $u_I=(\zero,\zero,2d,2e_I,\zero) \in \codi$ with $d\in\{0,1\}$
(resp. $u_J=(\zero,\zero,2d',2e_J,\zero)\in \codi$ with $d'\in\{0,1\}$).
Since $2v_{j_1}*v_{j_2} \not\in \codi$ (resp. $2v_{j_3}*v_{j_4}
\not\in \codi$) we have
$2v_{j_1}*v_{j_2}=u_I+(\zero,\zero,2,\zero,\zero)$ (resp.
$2v_{j_3}*v_{j_4}=u_J+(\zero,\zero,2,\zero,\zero)$). Therefore,
$2v_{j_1}*v_{j_2} + 2v_{j_3}*v_{j_4}=u_I+u_J\in\codi$

By Proposition \ref{lemm:cosetsKernel}, there exist $\bar{k}$ row vectors
$v_1,v_2,\ldots,v_{\bar{k}}$ in $\cG_S$, such that $\Phi(v_I) \not\in
K(C)$ for any nonempty subset $I\subseteq \{1,\ldots,\bar{k}\}$ and
$ker(C)= \gamma+2\delta-\bar{k}$.
Assume $\bar{k}$ is odd. We will show that there exists a subset
$I\subseteq \{1,\ldots,\bar{k}\}$ such that $\Phi(v_I)\in
K(C)$. Since this is a contradiction, $\bar{k}$ can not be an odd number
and the assertion will be proved.

By Lemma~\ref{lem3}, in order to prove that there exists $I\subseteq
\{1,\ldots,\bar{k}\}$ such that $\Phi(v_I)\in K(C)$, it is enough to
prove that $2v_I*v_j\in \codi$ for all $j\in\{1,\ldots,\bar{k}\}$.
That is, $2v_i*v_j\in \codi$ for all $i\in I$ and
$j\in\{1,\ldots,\bar{k}\}$ or, following the above remark, for each
$j\in\{1,\ldots, \bar{k}\}$ the number of $i\in I$ such that
$2v_i*v_j\not\in \codi$ is even. We define a symmetric matrix $A=(a_{ij})$, $1\leq
i,j\leq \bar{k}$, in the following way: $a_{ij}=1$ if $2v_i*v_j
\not\in \codi$ and $0$ otherwise. Therefore, $A$ is a symmetric matrix of odd order
and with zeroes in the main diagonal. Lemma
\ref{lemm:symmetric-matrix} shows that $\det(A)=0$ and hence there
exists a linear combination of some rows, $i_1,\ldots,i_l$, of $A$
equal to $\zero$. The vector $\Phi(v_I)$, where
$I=\{i_1,\ldots,i_l\}$, belongs to $K(C)$. This completes the proof.
\end{demo}

\begin{ex} Continuing with Example \ref{ex-Z4rank}, the dimension of the kernel for a Hadamard $\Z_4$-linear code $H$ was computed in \cite{Hadam} and \cite{Kro01} and the dimension of the kernel for an extended 1-perfect $\Z_4$-linear code $C$ in \cite{ExtPerf}.
Specifically,
$$ker(H)=\left\{
 \begin{tabular}{l l l}
   $\gamma+\delta+1$ & if & $\delta \geq 3$ \\
   $\gamma+2\delta$ & if & $\delta=1,2$ \\
\end{tabular}
  \right.$$
and $$ker(C)=\left\{
 \begin{tabular}{l l l}
   $\bgamma+\bdelta+1$ & if & $\delta \geq 3$ \\
   $\bgamma+\bdelta+2$ & if & $\delta=2$ \\
   $\bgamma+\bdelta+t$ & if & $\delta=1$. \\
\end{tabular}
  \right.$$
\end{ex}

\begin{ex}
 Continuing with Example \ref{ex-Z2Z4rank}, the dimension of the kernel for a Hadamard $\add$-linear code $H$ was computed in \cite{Hadam}  and the dimension of the kernel for an extended 1-perfect $\add$-linear code $C$ in \cite{ExtPerf}.
Specifically,
$$ker(H)=\left\{
 \begin{tabular}{l l l}
   $\gamma+\delta$ & if & $\delta \geq 2$ \\
   $\gamma+2\delta$ & if & $\delta=0,1$ \\
\end{tabular}
  \right.$$
and $$ker(C)=\left\{
 \begin{tabular}{l l l}
   $\bgamma+\bdelta+1$ & if & $\delta \geq 1$ \\
   $\bgamma+2\bdelta$ & if & $\delta=0$. \\
\end{tabular}
  \right.$$ Note that the kernel dimension of the Hadamard $\add$-linear codes satisfies the lower bound.
\end{ex}

\begin{ex}
Let $\overline{QRM}(r,m)$ be the class of $\Z_4$-linear Reed-Muller
codes defined in \cite{QRM}, as in Example \ref{ex-QRMrank}. The
dimension of the kernel of any code $C\in \overline{QRM}(r,m)$ is
$$
ker(C)=\sum_{i=0}^r \binom{m}{i} +1 = \delta +1,
$$
except for $r=m$ (in this case, $C=\Z_2^{2^{m+1}}$), \cite{QRM}.

Therefore, $\Z_4$-linear Kerdock-like codes  and extended
$\Z_4$-linear Preparata-like codes of binary length $4^m$ have
dimension of the kernel $ker(K)=2m+1$ and $ker(P)=2^{2m-1}-2m+1$,
respectively \cite{QRM}, \cite{PrepKerd}.
\end{ex}

As in Section \ref{section:rank} for the rank, the next point to be solved here is how to
construct $\add$-linear codes with any dimension of the kernel in
the range of possibilities given by Proposition \ref{bounds-kernel}.

\begin{theo} \label{all-kernels} Let $\alpha,\beta,\gamma,
\delta,\kappa$ be integer numbers satisfying
(\ref{bounds-code}). Then, there exists a $\add$-linear code $C$ of
type $(\alpha,\beta;\gamma, \delta;\kappa)$ with $ker(C)=k$ for any
$$k \in \left\{
 \begin{tabular}{l l l}
   $\{ \gamma +\delta, \ldots, \gamma+2\delta-2,\gamma+2\delta \}$ & if & $s\geq 2$ \\
   $\{ \gamma +2(\delta-\lceil \frac{\delta-1}{2} \rceil), \ldots, \gamma+2(\delta-1),
  \gamma+2\delta \}$ & if & $s=1 $ \\
  $\{ \gamma +2\delta \}$    & if & $s=0, $\\
\end{tabular}
  \right.$$
  where $s=\beta-(\gamma-\kappa)-\delta.$
\end{theo}

\begin{demo} Let $\codi$ be a $\add$-additive code of type
$(\alpha,\beta;\gamma, \delta;\kappa)$ with generator matrix
$$\cG= \left ( \begin{array}{cc|ccc}
I_{\kappa} & T' & \zero & \zero & \zero\\
\zero & \zero & \zero & 2I_{\gamma-\kappa} & \zero\\
\hline \zero & S' & S_k & \zero & I_{\delta} \end{array} \right
),
$$ where $S_k$ is a matrix over $\Z_4$ of size $\delta \times s$, and
let $C=\Phi(\codi)$ be its corresponding $\add$-linear code.
Taking $S_k$ as the all-zero matrix over $\Z_4$, the code $C$ is a binary linear code, so $ker(C)=k=\gamma+2\delta$.

When $s=1$, for each ${\bar k} \in \{ 2,4,\ldots, 2\lceil \frac{\delta-1}{2} \rceil \}$ and $k=\gamma+2\delta-{\bar k}$, we can construct a matrix $S_k$ over $\Z_4$ of size $\delta \times 1$ with an even number of ones,
${\bar k}$, and zeroes elsewhere. In this case, $ker(C)=k=\gamma+2\delta-\bar k$, by the proof of Proposition \ref{bounds-kernel}.

Finally, when $s\geq 2$, for each $\bar k\in \{2,3,\ldots,\delta \}$ and $k=\gamma+2\delta-{\bar k}$,
we can construct a matrix $S_k$ over $\Z_4$ of size $\delta \times s$, such that
only in the last $\delta-\bar k$ row vectors all components are zero and, moreover,
in the first $\bar k$ coordinates of each column vector there are an even number of ones and zeros elsewhere.
In this case, by the same arguments as in the proof of Proposition \ref{bounds-kernel}, it is easy to prove that $ker(C)=k=\gamma+2\delta-\bar k$.
\end{demo}

\begin{ex}
By Proposition \ref{bounds-kernel},
we know that the possible dimensions of the kernel for $\add$-linear codes, $C$, of type $(\alpha,9;2,5;1)$
are $ker(C)=k \in \{12,10,9,8,7 \}$. For each possible $k$, we can construct a $\add$-linear code $C$
with $ker(C)=k$, taking the following generator matrix of $\codi=\Phi^{-1}(C)$:
$$\cG_S= \left ( \begin{array}{cc|ccc}
1 & T'    & \zero & 0 & \zero\\
0 & \zero  & \zero & 2 & \zero\\\hline
\zero& S' &  S_k & \zero & I_5 \end{array} \right ),$$
where
$S_{12}=( \zero )$ and $S_{10}$, $S_9$, $S_8$ and $S_7$ are constructed as follows:
$$
S_{10}=\left(
    \begin{array}{ccc}
      1 & 0 & 0 \\
      1 & 0 & 0 \\
      0 & 0 & 0 \\
      0 & 0 & 0 \\
      0 & 0 & 0 \\
    \end{array}
  \right),
S_9=\left(
    \begin{array}{ccc}
      1 & 0 & 0 \\
      1 & 1 & 0 \\
      0 & 1 & 0 \\
      0 & 0 & 0 \\
      0 & 0 & 0 \\
    \end{array}
  \right),
S_8=\left(
    \begin{array}{ccc}
      1 & 0 & 0 \\
      1 & 1 & 0 \\
      1 & 1 & 0 \\
      1 & 0 & 0 \\
      0 & 0 & 0 \\
    \end{array}
  \right), 
S_7 =\left(
    \begin{array}{ccc}
      1 & 0 & 0 \\
      1 & 1 & 0 \\
      1 & 1 & 0 \\
      1 & 1 & 0 \\
      0 & 1 & 0 \\
    \end{array}
  \right). $$
\end{ex}

\section{Pairs of rank and kernel dimension of $\add$-additive codes}\label{section:pairs}

In this section, once the dimension of the kernel is fixed, lower and upper bounds on the rank are established. We will show that there exists a $\add$-linear code $C$ of type
$(\alpha,\beta;\gamma, \delta;\kappa)$ with $r=rank(C)$ and $k=ker(C)$ for any
possible pair of values $(r,k)$.

\begin{lemm}\label{lemm:bound_r_k}
Let $\codi$ be a $\add$-additive code of type $(\alpha,\beta;\gamma,
\delta;\kappa)$ and let $C=\Phi(\codi)$ be the corresponding $\add$-linear code. If $rank(C)=\gamma+2\delta+\bar{r}$ and
$ker(C)=\gamma+2\delta-\bar{k}$, with $\bar{k}\geq 2$, then $$1\leq \bar{r} \leq \binom{\bar{k}}{2}.$$
\end{lemm}

\begin{demo}
There exist $\{u_i \}_{i=1}^\gamma$ and $\{v_j \}_{j=1}^\delta$ vectors of order two and four respectively, such that they generate the code $\codi$ and $C= \bigcup_{I \subseteq \{1,\ldots,{\bar k} \}} (K(C)+ \Phi(v_{I}))$  by Proposition \ref{lemm:cosetsKernel}. Note that $\Phi(v_j)\in K(C)$ if and only if $j\in\{\bar{k}+1,\dots,\delta\}$.

By Lemma \ref{lem3}, for all $j\in\{\bar{k}+1,\dots,\delta\}$ and $i\in\{1,\dots,\delta\}$, as $\Phi(v_j)\in K(C)$, $2v_j*v_i \in\codi$ and, consequently, $\Phi(2v_j*v_i)$ is a linear combination of $\{\Phi(u_i)\}_{i=1}^{\gamma}$ and $\{\Phi(2v_j)\}_{j=1}^{\delta}$. As a result, $\langle C \rangle$ is generated by $\{\Phi(u_i)\}_{i=1}^{\gamma}$, $\{\Phi(v_j),\Phi(2v_j)\}_{j=1}^{\delta}$ and $\{\Phi(2v_t*v_s)\}_{1\leq s<t\leq \bar{k}}$ and hence $\bar{r}\leq\binom{\bar{k}}{2}$, by Lemma \ref{lemm:RankSet}.

Finally, since $\bar k \geq 2$, the binary code $C$ is not linear and, therefore, $\bar r\geq 1$.
\end{demo}

Let $C$ be a $\add$-linear code with $ker(C)=\gamma+2\delta-\bar{k}$ and $rank(C)=\gamma+2\delta+\bar{r}$. Note that if $\bar{r}=0$ then, necessarily, $\bar{k}=0$ (and viceversa) and $C$ is a linear code. The next theorem will determine all possible pairs of rank and dimension of the kernel for nonlinear $\add$-linear codes.

\begin{prop} \label{prop:all-rk} Let $C$ be a $\add$-linear code of binary length $n=\alpha+2\beta$ and type $(\alpha,\beta;\gamma,\delta;\kappa)$ with $ker(C)=\gamma+2\delta-\bar{k}$ and $rank(C)=\gamma+2\delta+\bar{r}$. Then,
$$\left\{
\begin{tabular}{l l}
   $\bar{r}\in \{ 2,\ldots, \min (\beta-(\gamma-\kappa)-\delta, \; \binom{\bar{k}}{2} ) \}$, & if $\bar{k} \in \{3,5,\ldots, 2\lceil \frac{\delta-1}{2} \rceil +1\}$,\\
   $\bar{r}\in \{ 1,\ldots, \min (\beta-(\gamma-\kappa)-\delta, \; \binom{\bar{k}}{2} ) \}$, & if $\bar{k}\in \{2,4,\ldots, 2\lceil \frac{\delta-1}{2} \rceil \}$.
\end{tabular} \right.$$
\end{prop}

\begin{demo}
By Proposition \ref{bounds-rank}, $\bar{r}\in \{ 0,\ldots, \min (\beta-(\gamma-\kappa)-\delta, \; \binom{\delta}{2}) \}$. Moreover, by Lemma \ref{lemm:bound_r_k}, for a fixed $\bar{k}\geq 2$, $\bar{r}\leq \binom{\bar{k}}{2}$ and, therefore, if $\bar{k} \in \{2,\ldots, \delta\}$ then $\bar{r}\in \{ 1,\ldots, \min (\beta-(\gamma-\kappa)-\delta, \; \binom{\bar{k}}{2}) \}$.

In the case $\bar{r}=1$, $C$ is not linear and, by Lemma \ref{prop:all-k},  $ker(C)=\gamma+2\delta-\bar{k}$ where $\bar{k}\in\{2,\dots,\delta\}$. Moreover, there exist $\bar k$ row vectors $v_1,v_2,\dots,v_{\bar{k}}$ of order four in any generator matrix $\cG$ of $\codi$  such that $C= \bigcup_{I \subseteq \{1,\ldots,{\bar k} \}} (K(C)+ \Phi(v_{I}))$, by Proposition \ref{lemm:cosetsKernel}. We will see that if $\bar{r}=1$, then $\bar{k}$ is necessarily even. Assume $\bar{k}$ is odd. We will prove that there exist $I\subseteq\{1,\dots,\bar{k}\}$ such that $\Phi(v_I)\in K(C)$, that is, $2v_I*v_j\in\codi$ for all $j\in\{1,\dots\bar{k}\}$, which is a contradiction and, therefore, $\bar{k}$ is an even number.

As $rank(C)=\gamma+2\delta+1$, by Lemma \ref{lemm:RankSet}, for all $i,j\in\{1,\dots,\bar{k}\}$ either $2v_i*v_j\in\codi$ or $2v_i*v_j=2v\notin\codi$. If there exist $I\subseteq\{1,\dots,\bar{k}\}$ such that, for each $j\in\{1,\dots,\bar{k}\}$ the number of $i\in I$ verifying $2v_i*v_j=2v\notin\codi$ is even, then $2v_I*v_j\in\codi$. In order to prove that there exist such a set $I$, we define the symmetric matrix $A=(a_{ij})$, $1\leq i,j\leq \bar{k}$, as in the proof of Proposition \ref{bounds-kernel}, and we get the contradiction.
\end{demo}

\begin{theo} \label{theo:all-rk} Let $\alpha,\beta,\gamma,\delta,\kappa$ be integer numbers satisfying
(\ref{bounds-code}). Then, there exists a $\add$-linear code $C$ of
type $(\alpha,\beta;\gamma, \delta;\kappa)$ with $ker(C)=\gamma+2\delta-\bar{k}$ and $rank(C)=\gamma+2\delta+\bar{r}$ for any
$$\left\{
\begin{tabular}{l l}
   $\bar{r}\in \{ 2,\ldots, \min (\beta-(\gamma-\kappa)-\delta, \; \binom{\bar{k}}{2} ) \}$, & if $\bar{k} \in \{3,5,\ldots, 2\lceil \frac{\delta-1}{2} \rceil +1\}$,\\
   $\bar{r}\in \{ 1,\ldots, \min (\beta-(\gamma-\kappa)-\delta, \; \binom{\bar{k}}{2} ) \}$, & if $\bar{k}\in \{2,4,\ldots, 2\lceil \frac{\delta-1}{2} \rceil \}$.
\end{tabular} \right.$$
\end{theo}

\begin{demo}
Let $\codi$ be a $\add$-additive code of type $(\alpha,\beta;\gamma, \delta;\kappa)$ with generator matrix
$$\cG= \left ( \begin{array}{cc|ccc}
I_{\kappa} & T' & \zero & \zero & \zero\\
\zero & \zero & \zero & 2I_{\gamma-\kappa} & \zero\\
\hline \zero & S' & S_{r,k} & \zero & I_{\delta} \end{array} \right
),
$$
where $S_{r,k}$ is a matrix over $\Z_4$ of size $\delta
\times(\beta-(\gamma-\kappa)-\delta)$, and let $C=\Phi(\codi)$ be its
corresponding $\add$-linear code.

Let $e_k$, $1\leq k \leq \delta$, denote the column vector of length
$\delta$, with a one in the $k$th coordinate and zeroes elsewhere. For each $\bar{k}\in\{3,\dots,\delta\}$ and $\bar{r}\in \{ 2,\ldots, \min (\beta-(\gamma-\kappa)-\delta, \; \binom{\bar{k}}{2}) \}$, we can construct $S_{r,k}$ as a quaternary matrix where in one column there is the vector $e_1+\cdots+ e_{\bar{k}}$, in $\bar{r}-1$ columns there are $\bar{r}-1$ different column vectors $e_k+e_l$ of length $\delta$, $1\leq k<l\leq \bar{k}$, and in the remaining columns there is the all-zero column vector. It is easy to check that $ker(C)=\gamma+2\delta-\bar{k}$ and $rank(C)=\gamma+2\delta+\bar{r}$.

Finally, if $\bar{r}=1$, we can construct $S_{r,k}$ as a quaternary matrix of size $\delta \times(\beta-(\gamma-\kappa)-\delta)$ with $\bar{k}$ ones in one column and zeroes elsewhere, for each $\bar{k}\in \{ 2,4,\ldots, 2\lceil \frac{\delta-1}{2} \rceil \}$. In this case, it is also easy to check that $rank(C)=\gamma+2\delta+1$ and  $ker(C)=\gamma+2\delta-\bar{k}$, for any $\bar{k}\in \{ 2,4,\ldots, 2\lceil \frac{\delta-1}{2} \rceil \}$.
\end{demo}

\begin{ex}
By Proposition \ref{prop:all-rk},
we know that the possible pairs of rank and dimension of the kernel of $\add$-linear codes, $C$, of type $(\alpha,9;2,5;1)$ are given in the following table:
\begin{center}
\begin{tabular}{c|cccc}
 $k \setminus r$ & 12 & 13 & 14 & 15  \\
 \hline
12&  * &   &   &     \\
10&    & * &   &     \\
9&    &   & * & *   \\
8&    & * & * & *   \\
7&    &   & * & *   \\
\end{tabular}
\end{center}
By Theorem \ref{theo:all-rk}, for each possible pair $(r,k)$, we can construct a $\add$-linear code $C$
with $rank(C)=r$ and $ker(C)=k$, taking the following generator matrix of $\codi=\Phi^{-1}(C)$:
$$\cG_S= \left ( \begin{array}{cc|ccc}
1 & T'    & \zero & 0 & \zero\\
0 & \zero  & \zero & 2 & \zero\\\hline
\zero& S' &  S_{r,k} & \zero & I_5 \end{array} \right ),$$
where $S_{12,12}=( \zero )$ and the other possible $S_{r,k}$ are constructed as follows:
$$
S_{13,10}=\left(
    \begin{array}{ccc}
      1 & 0 & 0 \\
      1 & 0 & 0 \\
      0 & 0 & 0 \\
      0 & 0 & 0 \\
      0 & 0 & 0 \\
    \end{array}
  \right),
 \quad
S_{13,8}=\left(
    \begin{array}{ccc}
      1 & 0 & 0 \\
      1 & 0 & 0 \\
      1 & 0 & 0 \\
      1 & 0 & 0 \\
      0 & 0 & 0 \\
    \end{array}
  \right),
$$
$$
S_{14,9}=\left(
    \begin{array}{ccc}
      1 & 1 & 0 \\
      1 & 1 & 0 \\
      1 & 0 & 0 \\
      0 & 0 & 0 \\
      0 & 0 & 0 \\
    \end{array}
  \right), \quad
S_{14,8}=\left(
    \begin{array}{ccc}
      1 & 1 & 0 \\
      1 & 1 & 0 \\
      1 & 0 & 0 \\
      1 & 0 & 0 \\
      0 & 0 & 0 \\
    \end{array}
  \right), \quad
S_{14,7} =\left(
    \begin{array}{ccc}
      1 & 1 & 0 \\
      1 & 1 & 0 \\
      1 & 0 & 0 \\
      1 & 0 & 0 \\
      1 & 0 & 0 \\
    \end{array}
  \right), $$
$$
S_{15,9}=\left(
    \begin{array}{ccc}
      1 & 1 & 0 \\
      1 & 1 & 1 \\
      1 & 0 & 1 \\
      0 & 0 & 0 \\
      0 & 0 & 0 \\
    \end{array}
  \right), \quad
S_{15,8}=\left(
    \begin{array}{ccc}
      1 & 1 & 0 \\
      1 & 1 & 1 \\
      1 & 0 & 1 \\
      1 & 0 & 0 \\
      0 & 0 & 0 \\
    \end{array}
  \right), \quad
S_{15,7} =\left(
    \begin{array}{ccc}
      1 & 1 & 0 \\
      1 & 1 & 1 \\
      1 & 0 & 1 \\
      1 & 0 & 0 \\
      1 & 0 & 0 \\
    \end{array}
  \right). $$
\end{ex}

\begin{ex} Again, by Proposition \ref{prop:all-rk},
the possible pairs of rank and dimension of the kernel of $\add$-linear codes, $C$, of type $(\alpha,18;2,6;1)$ are given in the following table:
\begin{center}
\begin{tabular}{c|cccccccccccc}
 $k \setminus r$ & 14 & 15 & 16 & 17 & 18 & 19 & 20 & 21 & 22 & 23 & 24 & 25\\
 \hline
14&  * &   &   &   &   &   &   &   &   &   &   &   \\
12&    & * &   &   &   &   &   &   &   &   &   &   \\
11&    &   & * & * &   &   &   &   &   &   &   &   \\
10&    & * & * & * & * & * & * &   &   &   &   &   \\
9&    &   & * & * & * & * & * & * & * & * & * &   \\
8&    & * & * & * & * & * & * & * & * & * & * & * \\
\end{tabular}
\end{center}

\end{ex}

\section{Conclusion}\label{section:conclusions}

In this paper we studied two structural properties of $\Z_2\Z_4$-linear codes,
the rank and dimension of the kernel. Using combinatorial enumeration techniques,
we established lower and upper bounds for the possible values of these parameters.
We also gave the construction of a $\Z_2\Z_4$-linear code with rank $r$
(resp. kernel dimension $k$) for each feasible value $r$ (resp. $k$).
Finally, we established the bounds on the rank, once the dimension of the kernel
is fixed, and we gave the construction of a
$\Z_2\Z_4$-linear code with rank $r$ and kernel dimension $k$ for each possible pair $(r,k)$.

The rank, kernel and dimension of the kernel are defined for binary codes and
they are specially useful for binary nonlinear codes.
We showed that for binary codes which are $\add$-linear codes,
we can also define the kernel using the corresponding $\add$-additive codes,
which are subgroups of $\Z_2^{\alpha}\times\Z_4^{\beta}$.
In this case, in order to compute the kernel $K(C)$ of a $\add$-linear code $C$ is much easier
if we consider the corresponding $\add$-additive code $\codi=\Phi^{-1}(C)$ and we compute
$\K(\codi)=\Phi^{-1}(K(C))$ using a generator matrix of $\codi$.
Moreover, we also proved that if $C$ is a $\add$-linear code,
then $K(C)$ and $\langle C \rangle$ are also $\add$-linear codes.
Finally, since $K(C)\subseteq C \subseteq \langle C \rangle$ and
$C$ can be written as the union of cosets of
$K(C)$, we also have that, equivalently, $\K(\codi) \subseteq \codi \subseteq  \Span_{\codi}$, where
$\Span_{\codi}=\Phi^{-1}( \langle C \rangle)$, and $\codi$ can be written as cosets of $\K(\codi)$.

As a future research in this issue, it would be interesting to establish a
characterization of all $\add$-linear codes of type $(\alpha,\beta;\gamma,\delta;\kappa)$ with
rank $r$ and dimension of the kernel $k$, using the canonical generator matrices $\cG_S$ of the form
(\ref{eq:StandardForm}) and characterizing their submatrices $S_{r,k}$ over $\Z_4$ of size
$\delta\times(\beta-\gamma-\delta)$.

\end{document}